\documentclass[slac_one]{revtex4}
\usepackage{graphicx}
\usepackage{fancyhdr}
\begin{document}

\title{Elucidating Jet Energy Loss in Heavy Ion Collisions} 

%

\author{N.~Grau \textit{for the ATLAS Collaboration}}
\affiliation{Columbia University, Nevis Laboratories, Irvington, NY 10533, USA}

\begin{abstract}
Very soon the LHC will provide beams for heavy ion collisions at 
5.52 TeV/nucleon. This center-of-mass energy results in a large cross-section 
for producing high-$E_T$ ($>$ 50 GeV) jets that are distinct from the soft, 
underlying event. This brings with it the possibility of performing full jet 
reconstruction to directly study jet energy loss in the medium produced in 
heavy ion collisions. In this note, we present the current state of jet 
reconstruction performance studies in heavy ion events using the ATLAS 
detector. We also discuss the possibilities of energy loss measurements 
available with full jet reconstruction: single jet $R_{AA}$ and di-jet and 
$\gamma$-jet correlations.
\end{abstract}

\maketitle

\thispagestyle{fancy}

\section{INTRODUCTION\label{sec:intro}}
Over the last eight years, the Relativistic Heavy Ion Collider (RHIC) at 
Brookhaven National Laboratory (BNL) has been providing heavy ion collisions 
in the attempt to create the Quark-Gluon Plasma (QGP) within the laboratory. 
Some of the most striking data from the first few years at RHIC were results 
from the apparent interaction of hard scattered partons in the medium produced 
at RHIC. Single high-$p_T$ particles are suppressed relative to binary 
collision-scaled $p+p$ rates\cite{PHENIXpi0s}; di-jets, measured from pairs of 
high-$p_T$ hadrons, are suppressed relative to $p+p$ and have a strongly 
modified angular correlation\cite{PHENIX:2008cq}; and heavy quarks, both charm 
and bottom, are suppressed as well\cite{Adare:2006nq}. These data have been 
interpreted as the loss of energy of a colored parton traversing a colored 
medium, analogous to energy loss in QED. However, many of the details 
underlying QCD energy loss are not well understood. On the one hand, energy 
loss models are able to reproduce the single particle suppression with very 
different assumptions on the rate of energy loss\cite{Renk:2006pw}. On the 
other, multiple high-$p_T$ particle correlations may suffer from a bias for 
the jet to lose little if any energy\cite{RenkPuncthrough,TangentialJets}. 
This is observed in data where the measured jet properties are quite similar 
between $p+p$ and $A+A$, and they have only slightly more constraining power 
on models.

To overcome such biases, it is necessary to move beyond multi-particle 
correlations and measure jets directly in heavy ion collisions. This note 
highlights the current status of jet reconstruction and energy loss 
measurements expected to be made with the ATLAS detector at the Large Hadron 
Collider (LHC). The large acceptance and nearly hermetic electromagnetic and 
hadronic calorimeters were designed for jet measurements and are uniquely 
suited to perform these measurements in heavy ion collisions. Such measurments 
will result in an increased understanding of how colored partons lose energy 
in a colored environment, a direct test of QCD.

\begin{figure*}[t]
\centering
\includegraphics[width=0.45\linewidth]{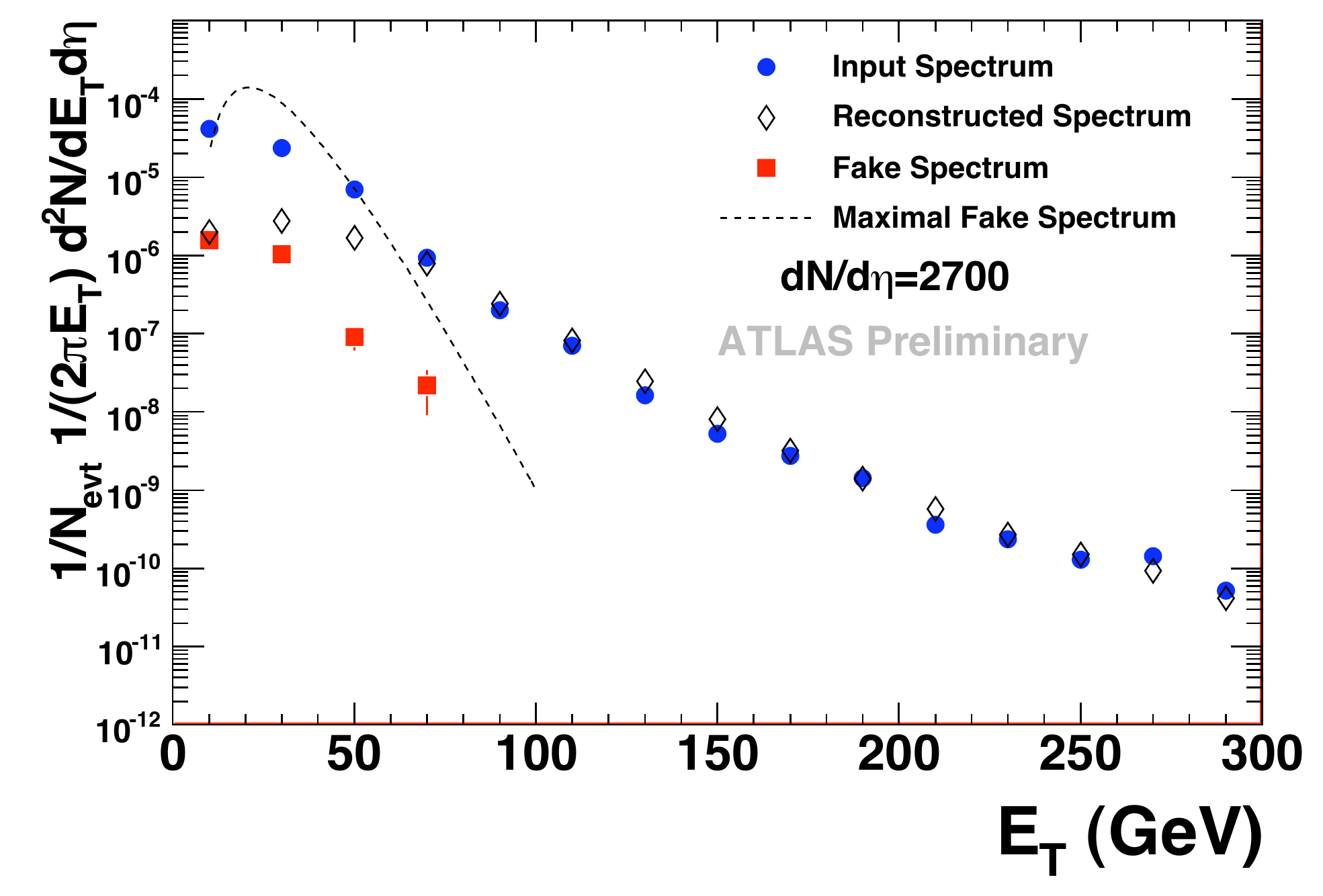}
\includegraphics[width=0.45\linewidth]{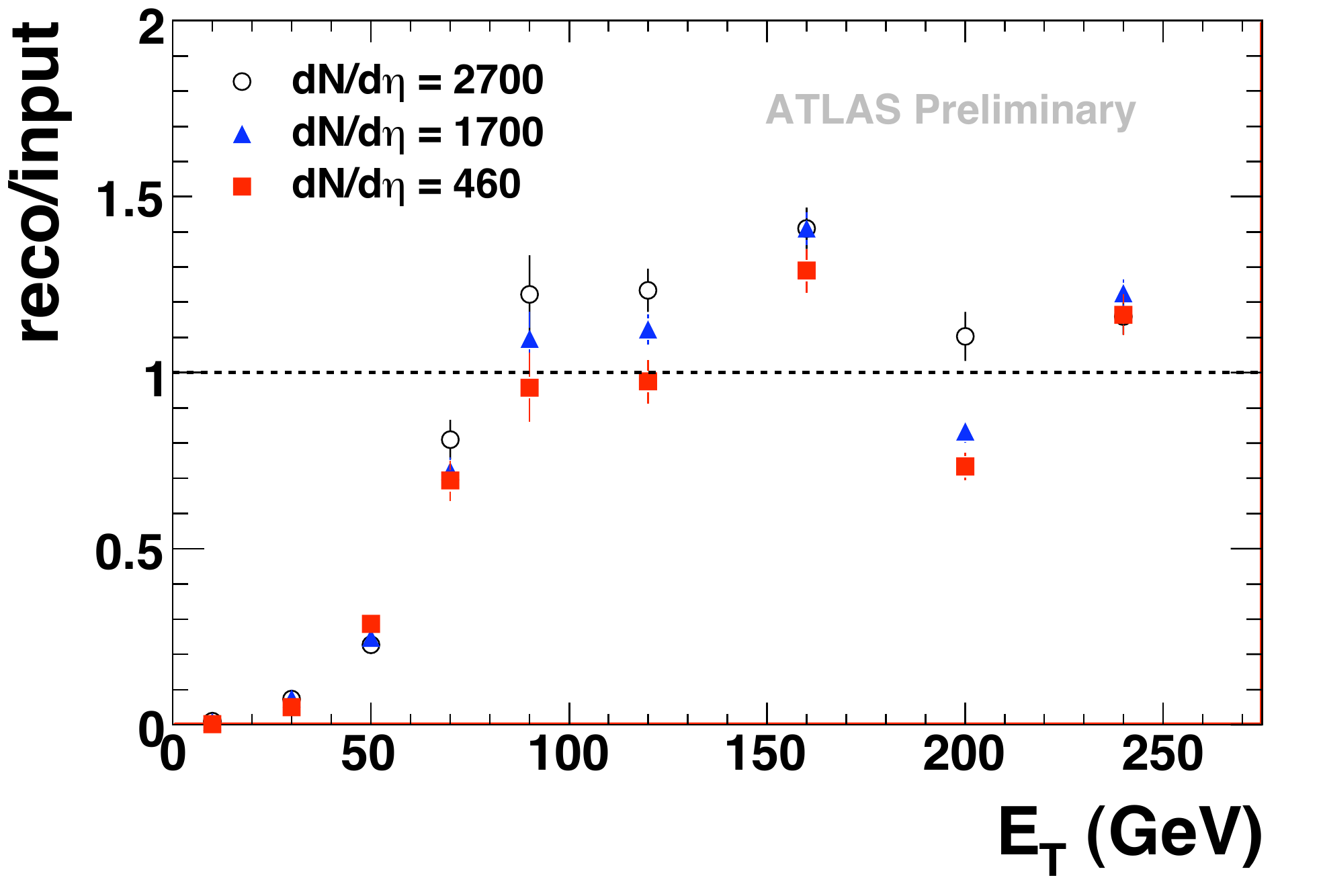}
\caption{
\textit{Left:} Binary-scaled PYTHIA jet spectrum (blue circles) compared to 
the raw reconstructed spectrum (open diamonds), the maximal fake rate (dashed 
line), and the residual fake rate (red squares). The raw reconstructed 
spectrum is uncorrected for efficiency and $E_T$ resolution.\textit{Right:} 
Ratio of raw reconstructed spectrum to the binary scaled expected spectrum for 
three different underlying event multiplicities. The 20\% difference at $E_{T} 
> $ 70 GeV before corrections indicates the level of sensitivity expected for 
jet $R_{AA}$ measurements. The non-statistical fluctuations in the data result 
from combining data sets from different $E_{T}$ ranges.
}
\label{fig:spectrum}
\end{figure*}

\section{JET RECONSTRUCTION\label{sec:jetreco}}
Unlike $e^{+}e^{-}$ annihilation or $p+p(\bar{p})$ collisions, heavy ion 
collisions result in a large, soft, underlying event. The charge particle 
multiplicity in central (head-on) $Pb+Pb$ collisions is expected to be between 
dN/d$\eta$=1500-3000. To put this in perspective, an average of 70-150 GeV 
would exist in an R = $\sqrt{\Delta\phi^{2}+\Delta\eta^{2}}$ = 0.4 radius cone 
from the underlying event alone. Therefore, jet clustering algorithms must be 
modified to handle such large backgrounds. Such a modification is not unlike 
the pile-up event subtraction\cite{Cacciari:2007fd} necessary at full LHC 
luminosity.

Both a seeded cone algorithm and the $k_T$ algorithm have been explored for 
use in heavy ion collisions within ATLAS\cite{Grau:2008ef}. For the rest of 
this note we will focus on seeded cone jet results. The underlying event 
energy is removed by the following steps. First, regions of high energy in the 
calorimeter are identified as possible regions with a jet. Next, the average 
$E_{T}$ from the cells are calculated as a function of $\eta$ and for each 
longitudinal calorimeter segment, excluding the regions in the first step. 
This average is then subtracted from {\it all} calorimeter cells. Finally, 
0.1$\times$0.1 towers in $\Delta\eta\times\Delta\phi$ built from a sum of 
subtracted cells are used as input to an R=0.4 cone algorithm with a tower 
threshold of $E_T >$ 5 GeV.

The left panel of Fig.~\ref{fig:spectrum} shows the comparison of the input, 
reconstructed, maximal, and final fake rate.several spectra. The input and 
reconstructed distributions are from di-jet PYTHIA events embedded in 
unquenched HIJING with dN/d$\eta$=2700. The raw reconstructed spectrum is not 
corrected for efficiency and $E_T$ resolutions. The raw fake spectrum is 
evaluated by running the jet algorithm with background subtraction directly on 
unquenched HIJING with a requirement that no jets with $E_T>$10 GeV are 
produced. The final fake rate is determined after a cut on the shape of the 
energy distribution within the jet has been made. Such a cut is also applied 
to the raw spectrum. For 70 GeV jets, the efficiency is 70\%, the energy 
resolution is 25\%, and the fake fraction is 3\%.

\section{MEASURING JET ENERGY LOSS\label{sec:eloss}}
Once jets can be reliably measured, their modification due to interactions 
with the medium will be explored. In this section we discuss modification of 
single jet rates\cite{Vitev:2008jh} and the increased acoplanarity of di-jets 
because of the incoherent multiple scattering in the 
medium\cite{DijetBroadening}.

The right panel of Fig.~\ref{fig:spectrum} shows the ratio of raw 
reconstructed to input spectra. Since the raw spectrum is uncorrected for 
efficiency and $E_T$ resolution, this represents a worst case scenario in the 
measurement of the jet $R_{AA}$ defined as 
\begin{equation}\label{eq:raa}
R_{AA} = \frac{\mathrm{Yield}_{A+A}}{\langle N_{coll} \rangle 
\mathrm{Yield}_{p+p}}
\end{equation}
Such a variable will be sensitive to details of energy loss. For example, if 
jets were perfectly reconstructed, $R_{AA}$=1. However, because collisional 
energy loss will impart energy to the medium, all lost energy will not be 
recovered and $R_{AA}<$1. Further, energy loss due to radiation outside of the 
cone size or the jet area will cause $R_{AA}<$ 1 of jets. A non-perturbative 
energy loss scenario based on AdS/CFT arguments would result in an 
$R_{AA}\ll$1\cite{Kharzeev:2008qr}. Therefore, studying the jet $R_{AA}$ as a 
function of cone size, seed $E_T$, etc. should be sensitive to the details of 
jet energy loss\cite{Vitev:2008jh}.

\begin{figure*}[t]
\centering
\includegraphics[width=0.45\linewidth]{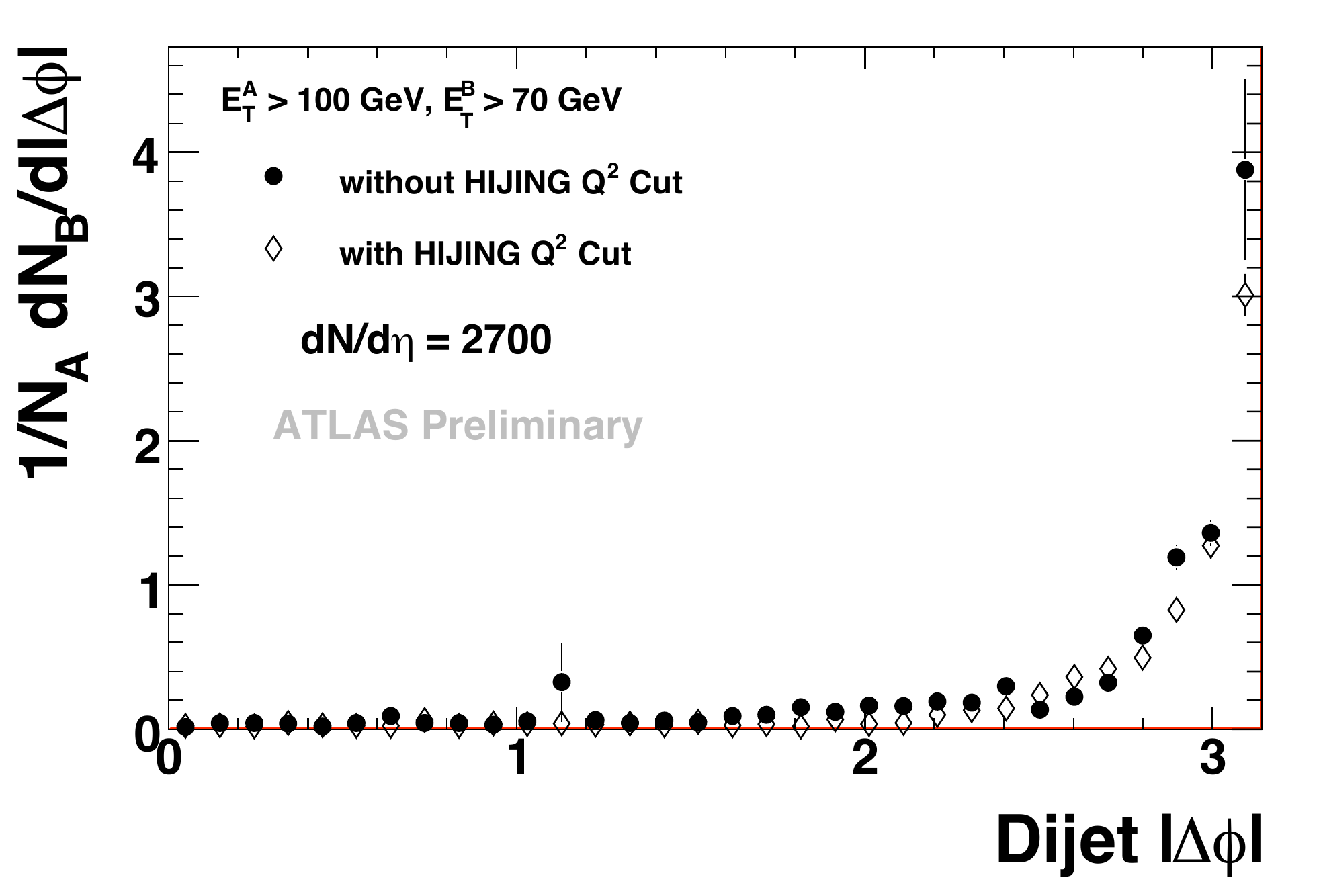}
\includegraphics[width=0.45\linewidth]{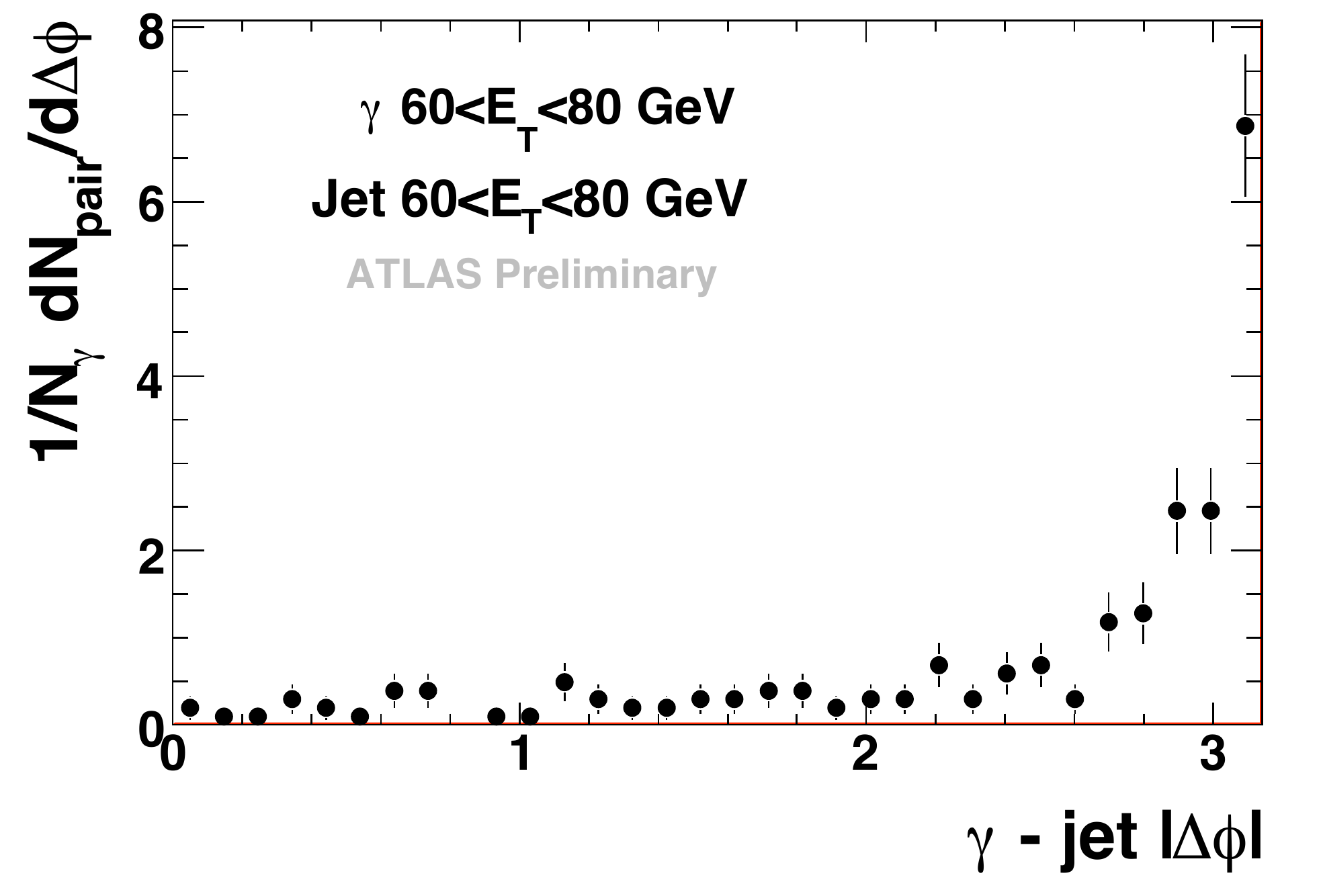}
\caption{
\textit{Left:} Correlations between pairs of reconstructed jets for ``trigger 
jet'' A $>$ 100 GeV and ``associated jet'' B $>$ 70 GeV. \textit{Right:} 
Correlations between reconstructed isolated photons and reconstructed jets for 
both the  $\gamma $ and jet from 60-80 GeV. The clear peaks at 
$\Delta\phi=\pi$ are indicative of $2\rightarrow2$ hard scattering. Multiple 
scattering and incoherent energy loss is expected to broaden these 
disributions\cite{DijetBroadening}. 
}
\label{fig:jetcorr}
\end{figure*}

Fig.~\ref{fig:jetcorr} shows two examples of jet azimuthal decorrelations to 
be measured in ATLAS. The left panel shows the reconstructed di-jet and the 
right panel shows the reconstructed $\gamma$-jet $|\Delta\phi|$ distributions. 
In a $2\rightarrow 2$ process, the produced hard scattering products are 
directly back-to-back, \textit{i.e.} at $|\Delta\phi|=\pi$. Intial and final 
state radition results in a natural broadening to this distributions. Further, 
radiation from in-medium energy loss will also broaden this 
distribution\cite{DijetBroadening}. However, this broadening has not been 
measured at RHIC\cite{STARjets}. This could indicate that the broadening is 
small or it is a consequence of the bias to little energy loss of the 
two-particle correlation measurements.

\section{CONCLUSION}
ATLAS expects to perform ground breaking measurements in QCD energy loss 
studies using fully reconstructed jets. This note highlighted the current 
status of cone jet reconstruction and measurements of jet rates and jet 
correlations as sensitive tests of QCD energy loss.

\end{document}